\documentclass{article}

\usepackage{PRIMEarxiv}

\usepackage[utf8]{inputenc} 
\usepackage[T1]{fontenc}    
\usepackage{hyperref}       
\usepackage{url}            
\usepackage{booktabs}       
\usepackage{amsfonts}       
\usepackage{nicefrac}       
\usepackage{microtype}      
\usepackage{lipsum}
\usepackage{siunitx}
\usepackage{amsmath}
\usepackage{fancyhdr}       
\usepackage{graphicx}       
\graphicspath{{media/}}     

\title{Performance Test Methodology for Atmosphere-Breathing Electric Propulsion Intakes in an Atomic Oxygen Facility
\thanks{\textit{\underline{Citation}}: 
\textbf{Authors. Title. Pages.... DOI:000000/11111.}} 
}

\author{
  Alexander T. Cushen \\
  Department of Climate and Space Sciences and Engineering \\
  University of Michigan \\
  Michigan\\
   \And
  Vitor T. A. Oiko, Katharine L. Smith, Nicholas H. Crisp, Peter C.E. Roberts  \\
  Space Systems Engineering Research Group, School of Engineering, \\
  University of Manchester \\
  Manchester\\
  \texttt{kate.smith@manchester.ac.uk} \\
  \And
  Francesco Romano \\
  Swiss Plasma Center (SPC) \\
  École Polytechnique Fédérale de Lausanne (EPFL) \\
  Lausanne\\
  \And
  Konstantinos Papavramidis, Georg Herdrich \\
  Institute of Space Systems \\
  University of Stuttgart \\
  Stuttgart\\
}

\begin{document}
\maketitle

\begin{abstract}
The testing of atmosphere-breathing electric propulsion intakes is an important step in the development of functional propulsion systems which provide sustained drag compensation in very low Earth orbits. To make satellite operations more sustainable, it is necessary to develop new materials which withstand erosion, long-lasting propulsion systems to overcome drag, and tools that allow for ground-based testing. Among the tools to enable these innovations is the Rarefied Orbital Aerodynamics Research facility at the University of Manchester. Here, a description of the facility is provided together with two different methodologies for testing sub-scaled intake designs for atmosphere-breathing electric propulsion systems. The first methodology is based on measurements of the pressure difference between the two extremities of the intake, while the second uses a gas sensor to measure the collection efficiency of the intake. Direct Simulation Monte Carlo models have been used to assess the viability of the proposed testing methodologies. The results of this analysis indicate that either methodology or a combination of both can provide suitable measurements to assess the performance of future intake designs. 
\end{abstract}


\newpage
\section{Introduction}
Very Low Earth Orbit (VLEO) (100-450km) is an attractive target for small communication and sensing satellites given the lower launch costs and proximity to the surface. At these altitudes, however, Atomic Oxygen (AO) is the primary atmospheric constituent and it presents a significant barrier to sustained spacecraft operations. The average AO number density 300 km above the surface is $10^{9}$ \si{cm^{-3}} \cite{Visentine1989} and the average orbital velocity of the spacecraft in relation to the oxygen atoms is approximately $7.7$ km s$^{-1}$ \cite{Murad1996}. This results in particle kinetic energies of $4.9$ eV and fluxes on the order of $10^{15}$ \si{atom} cm$^{-2}$  s$^{-1}$ in the ram direction. The local AO density is highly variable, influenced by a number of factors including solar activity, latitude, and variation in Earth’s magnetic field \cite{Zhang2002}. The collisions between the oxygen atoms and the surfaces of the spacecraft lead to a series of physio-chemical interactions that result in surface degradation through oxidation and erosion \cite{Reddy1995,Banks2003,Banks2004,MoeMoeWallace1998,Chernik2009}. They are also responsible for producing a drag force that results in orbital decay and a reduction of the mission lifetime. 

These conditions necessitate the consistent use of propulsion to maintain a stable orbit. Given the limited propellant which can be stored on spacecraft, Atmosphere-Breathing Electric Propulsion (ABEP) is a promising solution for extending mission lifetimes in low orbit. An ABEP system uses an intake mounted on the front of the spacecraft to collect gas and direct it through an internal channel to an electric thruster, which then uses the collected gas as a propellant.

A richer understanding of spacecraft surface drag and erosion allows for the development of novel materials and coatings that are better suited to the aggressive environment of VLEO, in line with the objective of the DISCOVERER program \cite{Roberts2019}. For instance, improved knowledge of the scattering mechanisms of atomic oxygen on different surfaces allows for the geometry of the spacecraft to be tailored to their specific missions. To this end, studies investigating the reflective properties of AO on borosilicate glass for fluence control in flight experiments \cite{TAGAWASEIKYUMAEDAEtAl2006} and of AO on gold, highly oriented pyrolytic graphite, and silicon dioxide for gas concentration mass spectrometry analyses on spacecraft \cite{MurrayPilinskiSmollEtAl2017} have been conducted. 

These works have been performed in ground-based facilities. However, the majority of the atomic oxygen exposure facilities operational today are focused on investigating the induced erosion, which is a serious concern as its effects can damage different systems such as optics and electronics with the potential of compromising whole missions \cite{YokotaTagawaOhmae2003, BuczalaBrunsvoldMinton2006}. Identifying materials that are resistant to AO erosion has been a key interest of these facilities \cite{ReddySrinivasamurthyAgrawal1993, Packirisamy1995}. To expedite the effects of AO exposure, many have increased AO flux to levels that are orders of magnitude above those experienced by spacecraft in VLEO \cite{Shpilman2008}. This undoubtedly is a helpful capability that allows the observation of what equivalently long exposure would cause to the materials, nevertheless, this carries the risk of becoming unrepresentative of the actual gas-surface interactions experienced whilst on orbit. 

Describing the gas-surface interactions (GSI) is a complex task that involves a great number of variables and uncertainties affecting both the gas particles and the surface, such as composition, morphology, energy of the gas, flux intensity, and flow regime, with different models applied to describe the results observed in the laboratory tests \cite{Goodman1967,SomorjaiBrumbach1973,Goodman1977,Livadiotti2020}. To properly characterize the dynamics, reactions, and processes involved, a facility that provides an environment comparable to VLEO, with appropriate AO flux and energy distribution, is required. To this end, a novel experimental facility that is designed to investigate the gas-surface interactions between material and engineering samples and a beam of AO is being developed at the University of Manchester \cite{Oiko2020}. Among the possible studies to be carried out at this facility is the characterization of the GSI of different materials and the \textit{in situ} testing and calibration of ABEP intake designs. In this work, we outline two methodologies for assessing the performance of sub-scale intakes in this facility. The first method is based on monitoring the collected gases via pressure measurements, while the second method analyzes the collection efficiency of the intake, using quartz crystal microbalances (QCM) as mass monitors. Direct Simulation Monte Carlo (DSMC) modeling was used to conduct numerical tests for the pressure measurements, and the model results were also used to verify that the QCM erosion rate is within detectable ranges.

\section{The Rarefied Orbital Aerodynamics Research facility}

The Rarefied Orbital Aerodynamics Research (ROAR) facility is designed to investigate the gas-surface interactions between a beam of $\sim 4.5$ \si{eV} neutral oxygen atoms and samples of different coatings and materials. To guarantee that the interactions observed are in good agreement with those of the real environment, ROAR has a set of systems to ensure that experiments are performed in conditions similar to those of VLEO, \textit{i.e.}, free molecular flow regime and AO fluxes in the correct range of values. These are provided by an ultrahigh vacuum (UHV) system and the atomic oxygen source (AOS), respectively.

Figure 1 shows a schematic of ROAR and the previously mentioned systems together with the sample holder and sample exchange chamber, providing an overview of the facility and how its components are distributed within the main chamber.

\begin{figure}[hbt!]
\centering
\includegraphics[width=.7\textwidth]{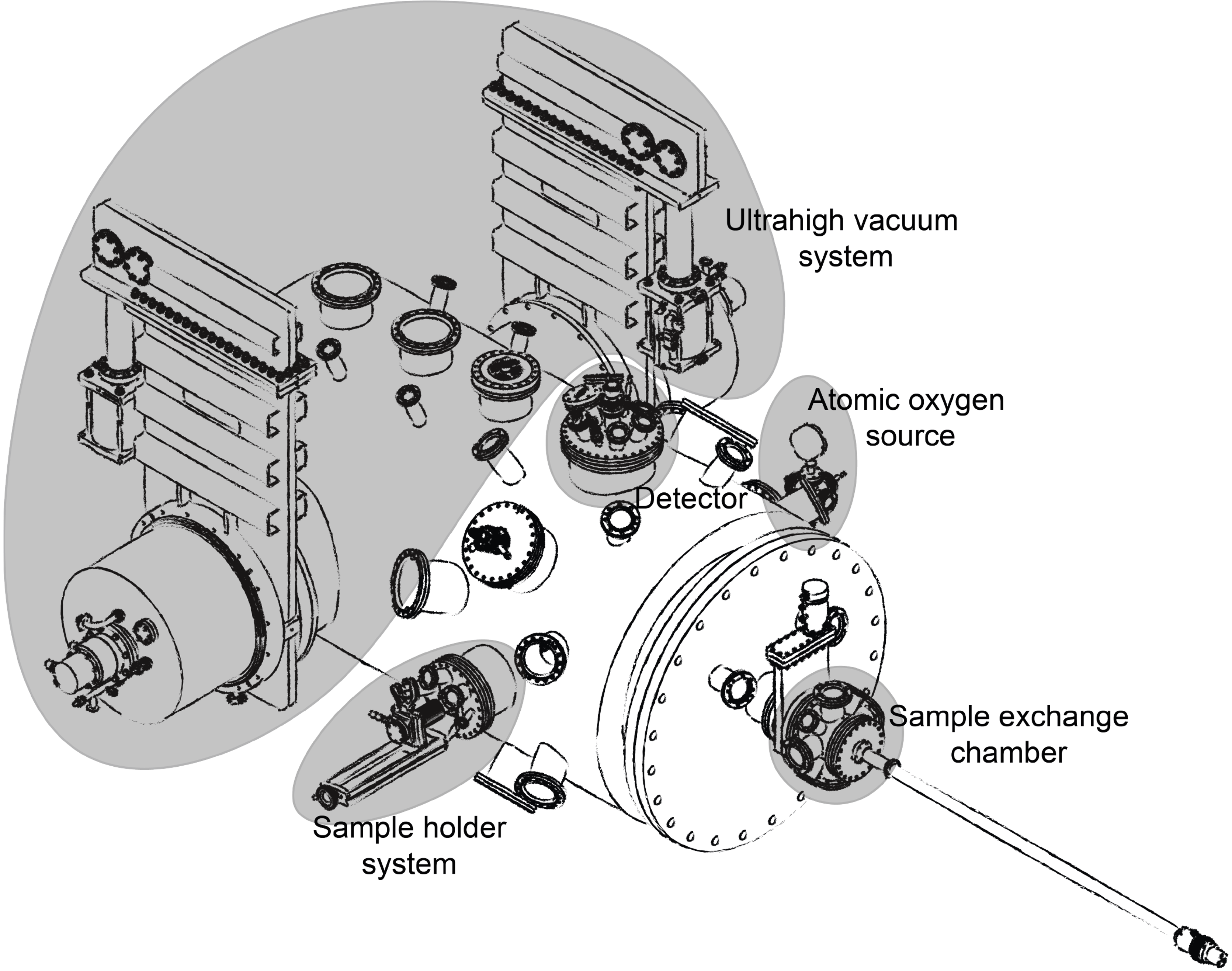}
\caption{Schematic of ROAR depicting the main systems described, including the ultrahigh vacuum system, atomic oxygen source, and detector.}
\label{fig:ROAR}
\end{figure}

\subsection{Ultrahigh vacuum system}
The ultrahigh vacuum system is composed of two cryopumps (Sumitomo CP-20) and a non-evaporable getter, NEG, (SAES, CapaciTorr D3500) that together provide $23.5$ \si{m^{3} s^{-1}} of pumping capacity. Working pressures are estimated to vary between $2\cdot10^{-7}$ and $2\cdot10^{-9}$ \si{hPa} depending on the emission profile of the atomic oxygen source, with the upper limit given to a cosine emission and the lower one to a collimated beam for a flux of $10^{15}$ \si{atom} \si{cm^{-2}} s$^{-1}$ measured at the sample. This range of pressures ensures that a free molecular flow regime is kept throughout the experiments, with mean free paths that are four orders of magnitude longer than the sample's dimension of $1$ \si{cm}, representing the system’s characteristic length. 

\subsection{Atomic oxygen source}
The atomic oxygen source (AOS) used in ROAR is based on the electron-stimulated desorption of oxygen atoms from a silver membrane \cite{Outlaw1987, Outlaw1994}. A silver membrane acts as an interface between two regions, a high-pressure side where O$_2$ is provided, and a low-pressure side maintained by the ultrahigh vacuum environment of the experiment’s chamber. 
The oxygen molecules adsorb onto the membrane and dissociate, before permeating through the membrane, reaching the UHV side where an electron beam impinging on the surface causes them to desorb, thus creating the beam of oxygen atoms. AO beams with kinetic energy of $4.5$-$5.0$ eV and a flux of $10^{13}$ atoms cm$^{-2}$ s$^{-1}$ has been achieved in previous experiments with this apparatus \cite{Outlaw1994}.

Increasing the AO flux to the expected level of $10^{15}$ \si{atom} \si{cm^{-2} s^{-1}} requires both the permeation and desorption processes to be optimized. Permeation can be increased through several measures: increasing the temperature of the silver membrane, decreasing its thickness, or by increasing the oxygen pressure on the high-pressure side. Desorption, on the other hand, can only be increased by applying higher electron currents. The source applied in ROAR has some modifications to the upstream side of the membrane that allows for a glow discharge to be formed on the high-pressure side of the membrane \cite{Outlaw1990,Wu1993,Premathilake2019}. 

The presence of a plasma causes both the temperature and the gas concentration to be increased near the surface, leading to an increase in permeation. For this improvement to be translated into AO flux, it must be accompanied by an increase in desorption, which is achieved by applying a high-current electron gun that provides between $5$-$50$ \si{mA} with electron energies varying between $0.1$-$10$ \si{keV} (Kimball Physics, EGG-3103). This configuration of the AOS is another innovative characteristic of ROAR. The independent control of both electron current and energy gives the possibility of isolating and analyzing the effects the electron energy has on the desorption of AO and consequently on the beam production. The AOS is mounted $12.5$ \si{cm} away from the sample and it is expected to have a beam around $1$ \si{cm} in diameter. 

\subsection{Detection system}

The detection system of ROAR is also another area where a novel approach has been taken. It comprises two different sensors: an Ion-Neutral Mass Spectrometer (INMS) and a residual gas analyzer (RGA, Pfeiffer, PrismaPro QMG250). The latter is a standard tool for monitoring the conditions of the vacuum inside the chamber, its cleanliness, and the composition of the residual atmosphere. For studying gas-surface interactions it is of fundamental importance to be able to monitor the conditions in which the experiments are performed. 

The INMS is a mass spectrometer built at the Mullard Space Science Laboratory, University College London that is composed of four sections: ion filter, ionizer, time-of-flight (TOF), and the analyzer \cite{Attrill2021}. The first two stages are used to select whether neutrals or ions are being detected, the TOF is used to determine the particle’s velocity, and the analyzer measures the particle’s energy. By mounting the INMS at different positions with respect to the sample and beam, the reflectance behavior of the sample can be ascertained. This provides the basis for analyzing materials and characterizing their orbital aerodynamic properties. It also serves as a testing platform for the development of sensors and payloads for VLEO.

For the study of intake designs for ABEP systems, auxiliary systems must be implemented to measure important parameters to assess the intake’s performance. The next sections consider the methodologies and the experimental systems required for measuring the intake’s collection efficiency $\eta_c$. 

\section{Testing platform for atmosphere-breathing electric propulsion systems}
ROAR has been designed to carry out tests of sub-scaled intake designs for ABEP systems. Within DISCOVERER, this research on the development of the intakes and the ABEP technologies is led by the Institute of Space Systems (IRS), University of Stuttgart \cite{Binder2016,Romano2018,Romano2020,Romano2021}. In this section, we will discuss the types of tests and experimental methodology that position ROAR to become an important characterization tool for propulsion systems and spacecraft surface properties.

\begin{figure}[hbt!]
\centering
\includegraphics[width=.9\textwidth]{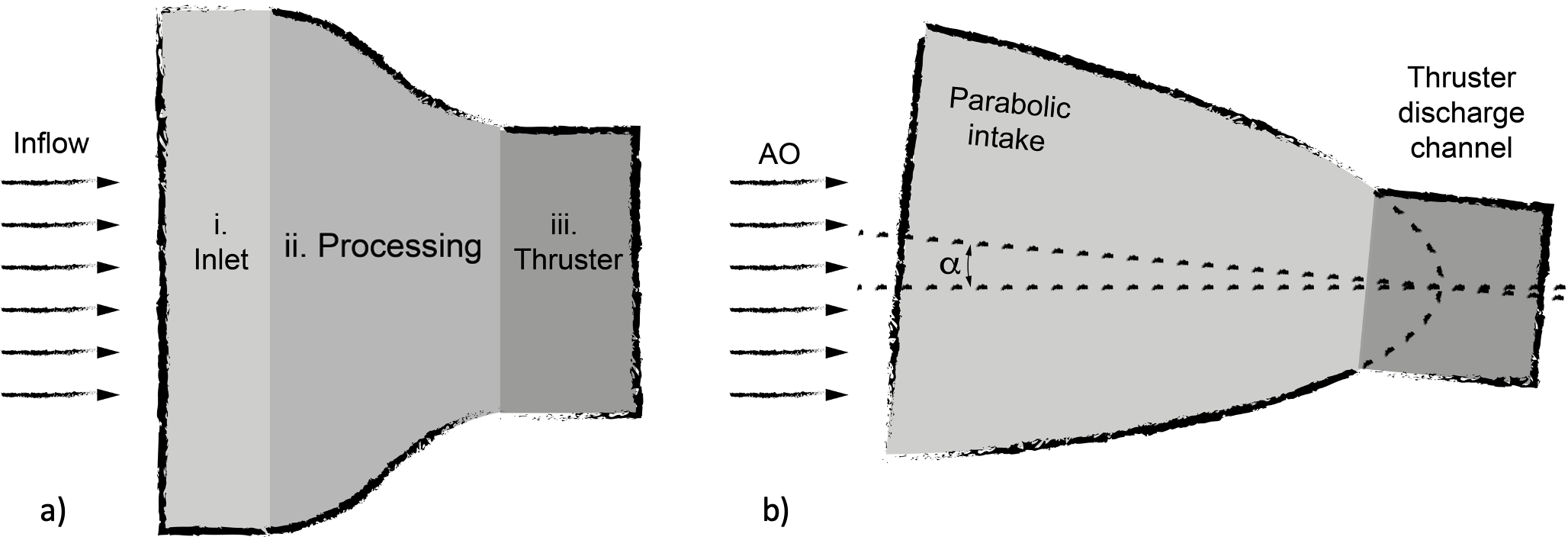}
\caption{Schematics of a) general propellant collection system and b) sub-scaled parabolic intake designed for tests at ROAR.} 
  \label{fig:PropCollection}
\end{figure}

A basic ABEP system is depicted on the left in fig. \ref{fig:PropCollection}. It can be divided into three main stages: an inlet responsible for collecting the oncoming flow which determines the amount of gas fed into the system, a second stage where handling of the collected particles varies according to the type of system used, \textit{i.e.}, the gas can be compressed, condensed, stored, etc, and finally the thruster itself. More detailed information on the design of such intakes can be found in \cite{Singh2015,Romano2020,Romano2021,Romano2022}. The intake characterization experiments to be conducted in ROAR focus on the performance of this first region, the inlet. The baseline conceptual design used in ROAR is shown on the right in fig. \ref{fig:PropCollection}.  It has been designed to minimize the impact from the downstream processing and thruster regions. It features a parabolic intake that directs particles into a thruster discharge channel, which delivers the collected particles into an auxiliary chamber where pressure measurements can be performed. In a real propulsion system, this discharge channel and the auxiliary chamber would be replaced by the processing and thruster regions. Parabolic intakes were chosen since they are known to have high collection efficiencies, collecting more than $92\%$ of incoming particles, assuming an ideal specularly reflecting surface \cite{Romano2021}. 

The sub-scaled intakes will be mounted on the same sample holder designed for the material testing, which provides the control of the angle of alignment between the AO beam and the intake, $\alpha$. A minimum step of $1.8^\circ$ and a precision of $5\%$ is achieved with a stepper motor (Arun Microelectronics, D35.1). Collection efficiency for a parabolic intake coated with materials that promote specular reflections is expected to vary non-linearly with $\alpha$, going from above $90\%$ of efficiency for $\alpha=0^{\circ}$ to around $10\%$ for $\alpha=20^{\circ}$ \cite{Romano2021}. Understanding the effects of misalignment on the intake's performance is of fundamental importance to the study of new intake designs as it can be translated into future requirements for the spacecraft's attitude control.

A key parameter to evaluate the intake performance is its collection efficiency, given by the ratio of the total number of particles collected to the total number of particles in the inflow. In ROAR this can be accomplished in two ways, by either measuring the pressure at the end of the intake or using a mass sensor. We detail the methodology for each technique and present experimental intake designs which suit the methods employed. These designs focus on reliably validating the intake performance and do not represent a highly-optimized or fully-integrated ABEP system. At present, ROAR will be used for benchmarking the real-world performance of intake inlets, without considering the impact of downstream processing and thruster systems.

\subsection{Differential pressure method of determining ABEP intake performance}
For the pressure measurements, the end of the intake must be connected to a second smaller auxiliary chamber with an independent vacuum system, pressure gauge, and a conductance control valve to regulate the pressure difference between the chambers. By knowing the system's dimensions such as volume, diameter, and length, it is possible to estimate the conductance between the two chambers and the increase in pressure that will be observed during the experiments. Since the incoming flow is characterized by ROAR's detection system, pressure readings can therefore be linked with the number of particles collected by the intake, thus providing the intake's collection efficiency. This approach can be supported by numerical analyses based on particle codes, see \cite{Munz2014,Fasoulas2019,Romano2021}. 

\begin{figure}[hbt!]
\centering
\includegraphics[width=.8\textwidth]{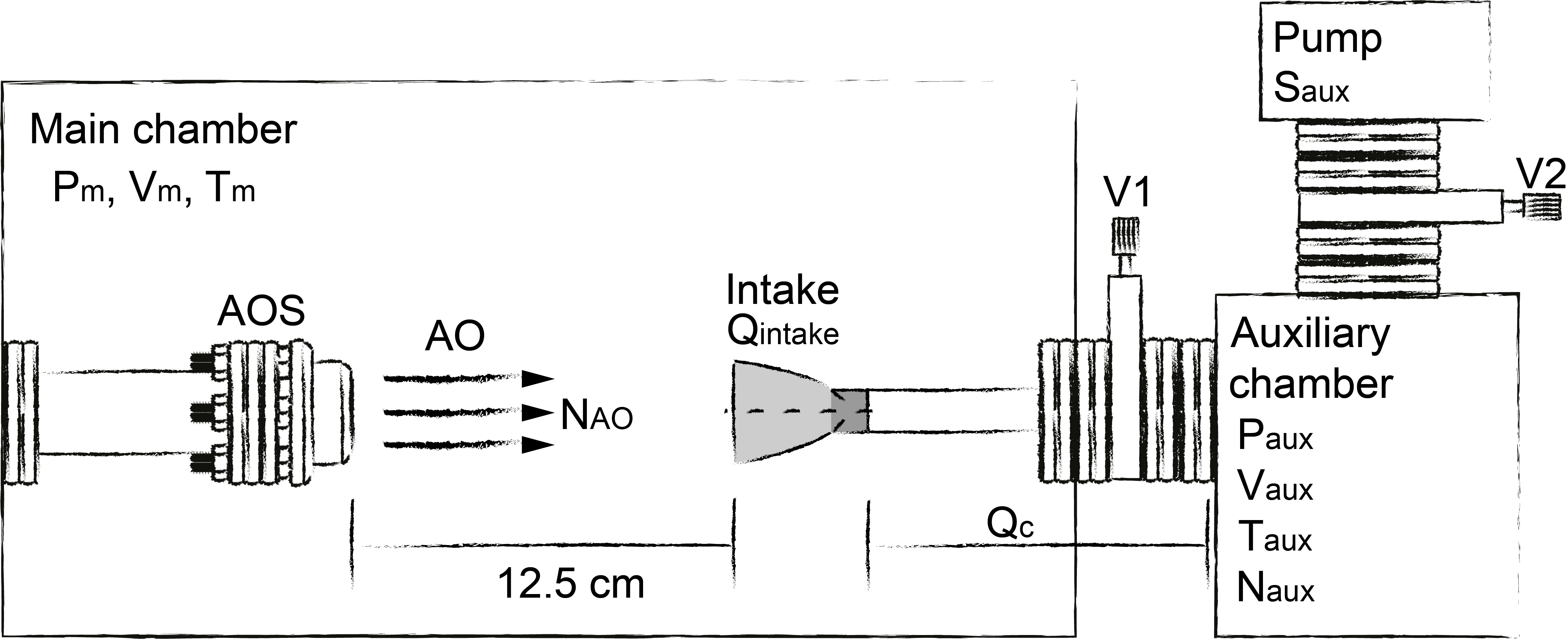}
\caption{Experimental setup for characterizing intake performance via pressure measurements, showing the atomic oxygen flux $N_{AO}$ being directed by intake into an auxiliary chamber.} 
\label{fig:PressureSetup}
\end{figure}

A schematic of the experimental setup for the pressure measurements is presented in fig. \ref{fig:PressureSetup}. It shows the main chamber with the atomic oxygen source (AOS) and the intake, the auxiliary chamber, the isolation valve (V$1$) and the connections between the chambers, a conductance control valve (V$2$), and the vacuum pump for the auxiliary chamber. In the experiments, the pressure, temperature, and volume for both chambers are known parameters; $p_{m}$, $T_{m}$, and $V_{m}$ for the main chamber, and $p_{aux}$, $T_{aux}$, and $V_{aux}$ for the auxiliary chamber. These are obtained from pressure gauges and temperature sensors installed in each chamber. The number of inflow particles at the intake, $N_{AO}$, is also measured by ROAR's detection system, specifically the INMS. The unknown quantities are the conductance of the intake, $Q_{intake}$, of the connections, $Q_{c}$, and the pumping speed, $S_{aux}$.

$S_{aux}$ can be determined by isolating the two chambers (V$1$ closed), adjusting V$2$, and comparing it with $p_{aux}$. The conductance control valve allows for the pumping speed to be varied so that the pressure difference between the chambers, and consequently at the back of the intake, can be adjusted, with maximum pumping speeds obtained when V$2$ is completely open. To measure the conductance of the connections between the chambers, $Q_{c}$, the atomic oxygen source is turned off, the isolation valve, V$1$ is opened, and the pressures $p_{m}$ and $p_{aux}$ are measured. Together with $S_{aux}$ previously determined and the other parameters of the system, temperature, and volume, $Q_{c}$ can be determined, leaving only $Q_{intake}$ as an unknown.

The intake's conductance can be determined by turning on the atomic oxygen source and monitoring how the pressures and temperatures in the two chambers vary. The conductance control valve can be adjusted such that the system is kept at a steady state condition so that $Q_{intake}$ can be measured. Because all the other parameters have already been defined, the number of particles in the back of the intake can be calculated, $N_{aux}$, as well as the collection efficiency, $\eta_c = N_{aux}/N_{AO}$. A similar approach has been considered and discussed for the testing of VLEO intakes in other facilities \cite{Brabants2021}.

\subsection{Verification of differential pressure method using numerical modeling}
To contextualize the results from differential pressure readings, we present the results of a set of numerical simulations which explore the impact of different experimental parameters on pressure-time values. These numerical experiments are conducted using Direct Simulation Monte Carlo (DSMC), a technique ideal for rarefied fluid environments that has been applied widely to study ABEP intake performance  \cite{Romano2021,Rapisarda2021,Parodi2019}. The DSMC simulations are conducted with the $dsmcFOAM+$ extension to the computational fluid dynamics platform $OpenFOAM$ \cite{White2018}.

First, we focus on simulating purely the front funnel region of the intake, composed of a parabolic section that collects and directs oncoming particles into the ABEP system. The intake model is produced in AutoCAD and converted into a high-resolution mesh file (approx. 80,000 cells) using SimFlow. The front of the mesh is the inlet patch that injects atomic oxygen DSMC particles into the domain at the density, velocity, and pressure specified and removes any backstreaming particles which reach it. By default, we set an inlet free-stream density of $1.4\times10^{14} \text{ m}^{-3}$. The outlet patch at the end of the domain also removes particles from the simulation. The walls of the intake funnel are set to have some probability of producing either specular or diffuse reflections when a particle collides with them. Specular-type reflections result in particles maintaining their temperature and speed, and deflecting off the surface at an angle equal to their incidence angle. Diffuse-type collisions scatter the particle into a direction independent of the incidence angle, effectively re-emitting the particle with a temperature and velocity set by a Maxwellian distribution. An intake temperature of $273$ K is adopted for this Maxwellian. Across our simulations, we vary the probability of reflection type to model the performance of different surface interaction regimes. 

Free molecular flow is assumed, meaning no interparticle collisions are considered. This assumption was validated by repeating many of the simulations we conducted with a hard sphere model for interparticle collisions, and confirming that no significant difference appears when these collisions are considered. In fact, no significant difference was found between collisionless and collisional simulations until the density was increased several orders of magnitude above the densities reached in a normal experiment. Each simulation is run until it converges, meaning the number of DSMC particles and mean particle kinetic energy in the domain reaches a steady state. The number of DSMC particles was chosen to be 10-20 times the number of mesh cells. Once convergence is reached, fluid properties like density and velocity are sampled in each cell for an extended period to minimize stochastic fluctuations of the values. The number of particles leaving through the outlet is then calculated by integrating the number flux across the outlet surface, and the incoming flux is computed analytically using eq. (2.19) in \cite{Parodi2019}. 

\begin{figure}[hbt!]
\centering
\includegraphics[width=.7\textwidth]{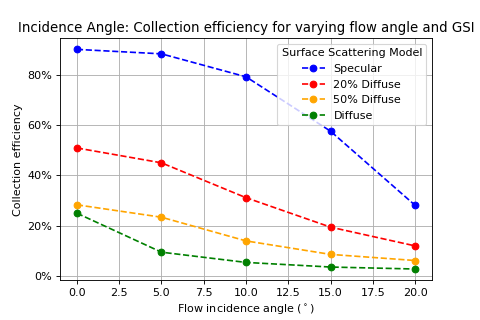}
\caption{\label{fig:incAngle} Collection efficiencies of subscale intake design for different incidence angles of the incoming flow. Color indicates surface scattering models used.}
\end{figure}

We first model the performance of the intake funnel over a range of incidence angles and surface interaction regimes. Figure \ref{fig:incAngle} shows the simulated collection efficiencies as the angle between the incoming flow and intake axis is increased from $0^\circ$ to $20^\circ$. This change in angle is achieved by decreasing the inlet patch velocity component parallel to the intake axis and increasing the perpendicular component. Similar to previous studies, we find an efficiency in excess of 90\% for low incidence angles for a perfectly specular parabolic intake, which decreases non-linearly for larger angles \cite{Romano2021} and more diffuse surface interaction regimes. 

\begin{figure}[hbt!]
\centering
\includegraphics[width=.7\textwidth]{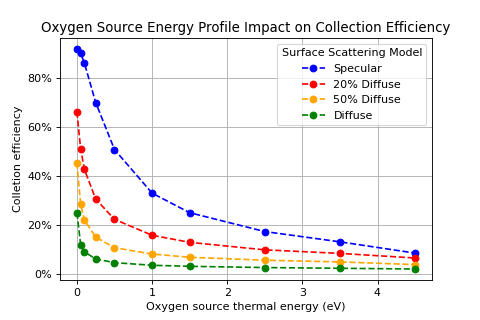}
\caption{Collection efficiencies for different AOS energy profiles. Color indicates surface scattering models used.}
\label{fig:AOS}
\end{figure}

Figure \ref{fig:AOS}  shows a similar comparison of $\eta_c$ for different energy distributions of the AO source. Using $4.5$ eV as a total energy baseline, we partition the energy between "thermal" energy, which introduces random thermal noise to the particle velocity, and "kinetic" energy, which sets the bulk velocity of the free-stream AO. As expected, we find that an AO source with lower thermal noise results in higher collection efficiencies. For the simulations which follow, we assume an AO source of ($T = 523$ K, $v = 7340$ m/s) and a purely specular surface in the parabolic region of the intake. Table \ref{table:input parameters} lists the standard input parameters in detail. These temperature and velocity values were adopted under the assumption almost all of the $\sim4.5$ eV delivered to each particle by the AOS is directed into linear kinetic energy but with a non-vanishing thermal noise. The true performance of ROAR's AOS will need to be assessed to validate this.

\begin{table}[hbt!]
\caption{\label{table:input parameters} Baseline input parameters used for \textit{OpenFOAM} simulations}
\centering
\begin{tabular}{l c}
\hline
\hline
Parameter & Value \\
\hline
Particle flux & $1.028 \times 10^{18}$ m$^{-2}$ s$^{-1}$ \\
Bulk velocity & $7340$ m s$^{-1}$  \\
AO temperature & $523$ K \\
Intake surface temperature & $273$ K \\
\hline
\end{tabular}
\end{table}

\begin{figure}[hbt!]
\centering
\includegraphics[width=.8\textwidth]{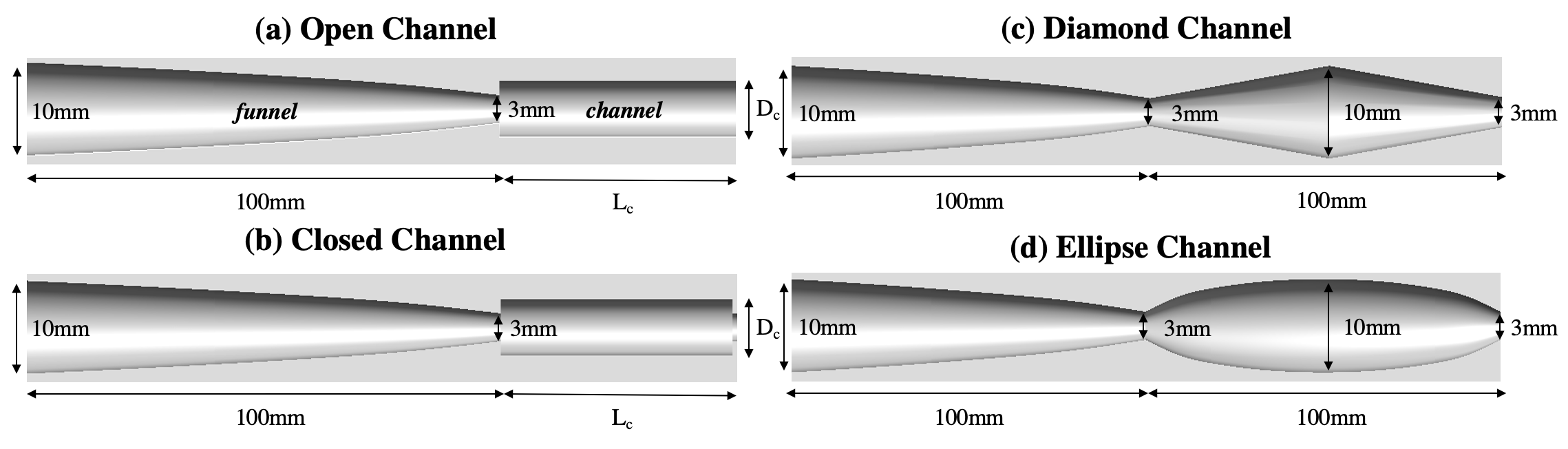}
\caption{Cross-sections of channel geometries adopted. Incoming flow is injected from the left side and the collection chamber is positioned to the right.} 
\label{fig:designs}
\end{figure}

To connect these results to observable values in ROAR, we then proceed to consider the impact
on collection efficiency from the channel which connects the intake to the "point-of-no-return";
an auxiliary chamber in ROAR, or the thruster in a complete ABEP system. The inclusion of this channel behind the funnel results in collection efficiencies for the entire system which are lower than those when only the funnel is considered but allows for a closer comparison between experimental results in ROAR and the performance of an entire intake assembly. Since the exact configuration of this channel system is yet to be designed, the 4 geometries we investigate, as shown in fig. \ref{fig:designs}, are chosen to span a range of possible experimental designs and provide some intuition for the impact of different geometric features. For instance, the Closed geometry introduces a constriction between the channel and the auxiliary chamber, allowing for a comparison between Open and Closed geometries which reveals the penalty of having a sharp barrier at the end of the channel. The Diamond and Ellipse channels demonstrate the impact of shape and curvature on efficiency, which may also feature in an ABEP propulsion system. In all cases, the channel walls are modeled as a diffuse surface, promoting only Maxwellian thermal-type reflections.

\begin{figure}[hbt!]
\centering
\includegraphics[width=.7\textwidth]{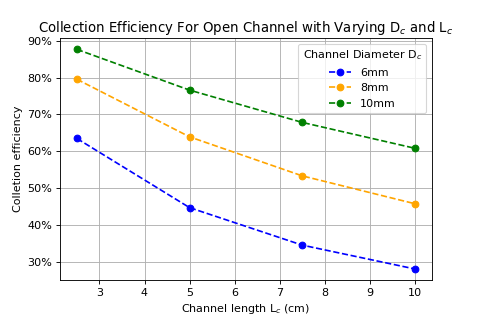}
\caption{\label{fig:dimensions} Collection efficiencies for different dimensions of ``open'' channel design, as a function of channel length. Color indicates channel radius.}
\end{figure}

First, we focus on the Open geometry as a baseline. Figure \ref{fig:dimensions} shows the simulated collection efficiency of the entire intake and channel system for this geometry, for a range of channel diameters $D_C$ and lengths $L_c$; in general, a wider, shorter channel results in higher efficiency. In the limiting case of a very short channel, the simulation reproduces the collection efficiencies observed when only the parabolic inlet is simulated.

In these simulations, particles that reach the end of the channel are removed from the simulation, and the population of particles that accumulate in the auxiliary chamber is not directly modeled due to the computational cost of doing so. In a real experiment in ROAR, however, this chamber will be "leaking" particles back through the channel and out the intake, reducing the pressure observed. Similar to the method proposed by Brabants \cite{Brabants2021}, we consider the rate of change of number density in the rear chamber $dn/dt$ of volume $V_{aux}$ to be the difference between the number flux coming in from the intake, $N_{in}$, and the flux backstreaming out of the chamber, $N_{out}$ given by eq. \ref{eqn:dn_dt}.

\begin{equation}
\label{eqn:dn_dt}
\frac{dn}{dt} = \frac{1}{V_{\text{aux}}}(N_\text{in} - N_\text{out}(t))
\end{equation}

By assuming the chamber is full of completely thermalized particles at temperature $T_{aux}$, and
the outlet from the channel to the chamber has an area $A_{aux}$, we can re-express as eq. \ref{eqn:Nout}.

\begin{equation}{\label{eqn:Nout}}
   N_\text{out}(t) = \eta_{rpc} A_{aux} \sqrt{\frac{k_B T_{aux}}{2 \pi m}} n(t) 
\end{equation}

where $k$ is Boltzmann’s constant and $m$ is the particle mass. We also introduce the reverse
performance coefficient $\eta_{rpc}$, which represents the fraction of thermalized particles leaking from the chamber which ultimately escape the inlet without being redirected back to the chamber. For instance, $\eta_{rpc} = 0$ would indicate that the intake assembly perfectly confines all thermalized gas in the auxiliary chamber, allowing the pressure to build indefinitely. In reality, this value is likely to be around $0.3$, which implies an equilibrium will be reached where the incoming and backstreaming fluxes are equal. By introducing this coefficient, we are assuming that all particles beyond the simulation domain are completely thermalized and that interparticle collisions throughout the intake can be neglected. By substituting $N_{in}$ = $\eta_c$ $N_{AO}$ and applying the ideal gas law, we get the expression for the chamber pressure $p_{aux}$ given by eq. \ref{eqn:PressureEq}.

\begin{equation}{\label{eqn:PressureEq}}
    p_{aux}(t) = \frac{\eta_c N_{AO}}{\eta_{rpc}A_{aux}}\sqrt{2 \pi m k_B T_{aux}}\left[1-\exp\left(  -\frac{\eta_{rpc}A_{aux}}{V_{aux}} \sqrt{\frac{k_B T_{aux}}{2 \pi m}} t\right) \right]
\end{equation}

Experimentally, the objective is to record chamber pressure over time and perform
an exponential fit to the data to determine the unknown parameters of interest, namely $\eta_c$. To ensure the pressure in the auxiliary chamber begins close to zero, the auxiliary pump $S_{aux}$ should be active until the experiment start time of $t=0$ s when valve $V_2$ is closed, allowing gas to accumulate. This approach can also be used to characterize AO beam parameters like $N_{aux}$ by repeating analogous experiments for different intake geometries (different $\eta_c$). Alternatively, to reduce the number of separate experiments performed, the system can be run with $S_{aux}$ active until steady state pressures are reached in the auxiliary chamber, before adjusting parameters like the intake incidence angle or AO source electron beam energy and monitoring the live response in the pressure.

\begin{figure}[hbt!]
\centering
\includegraphics[width=.7\textwidth]{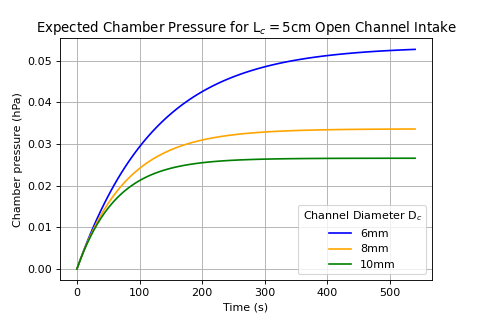}
\caption{\label{fig:pressure} Expected chamber pressure $p_{aux}$ measurements over time for different channel radii. }
\end{figure}

To estimate $\eta_{rpc}$ for our proposed intake geometries, we perform "reverse" simulations in which only the thermalized particles leaking from the chamber are modeled. The true flow in an experiment can then be inferred as the combination of the forward and reverse flows since the intake is operating in a collisionless regime. Table \ref{table:1} shows the collection efficiencies for each geometry considered, as well as the associated $\eta_{rpc}$. A smaller value for $\eta_{rpc}$ is generally favorable for an experiment since it allows for greater pressures to be reached in the auxiliary chamber. The elliptical channel was found to have the smallest $\eta_{rpc}$  of 3.9\%, but at the cost of lower collection efficiency. The open channel geometry was found to provide the best balance between high $\eta_c$ and low $\eta_{rpc}$, especially for wider diameter channels. To demonstrate these results, fig. \ref{fig:pressure} shows the expected auxiliary chamber pressure-time series for different channel radii using geometry (a) with chamber length $L_c$ = $5$ cm, and auxiliary chamber volume $V_{aux}$ = $0.05$ m$^3$. In each case, the pressure reached is within the detection range of the pressure sensors and maximizes over a timescale of minutes, which suits the capabilities of ROAR.

Longer channels tend to decrease the mass flow rate through the intake assembly since they provide a larger surface for diffuse collisions to scatter the particles and redirect them upstream. Closed-type channels also decrease the number of particles reaching the auxiliary chamber, as a consequence of the additional constriction to the flow they present. The competition between these effects can result in experimental designs with the same collection efficiency. Namely, a similar collection efficiency around $30\%$ can be attained for both a $L_c = 10.0$ cm, $D_c = 6.0$ mm Open channel design and a $L_c = 5.0$ cm, $D_c = 6.0$ mm Closed channel design. These designs do not share similar $\eta_\text{rpc}$ values, so they can nonetheless be distinguished by the rate of chamber pressure increase.

\begin{table}[hbt!]
\caption{\label{rpcTable} Summary of collection efficiencies ($\eta_c$) and reverse performance coefficients ($\eta_\text{rpc}$) determined from simulations.}
\centering
\begin{tabular}{c c c c c c}
\hline
\hline
Channel Type & $L_c$ (cm) & $D_c$ (mm) & $N_\text{out}$ (1/s) & $\eta_c$ & $\eta_\text{rpc}$\\
\hline
(a) Open & 5.0 & 6.0 & 3.61E+13 & 44.7\% & 9.4\% \\
(a) Open & 5.0 & 8.0 & 5.16E+13 & 63.9\% & 8.4\% \\
(a) Open & 5.0 & 10.0 & 6.18E+13 & 76.6\% & 6.8\% \\
(a) Open & 10.0 & 6.0 & 2.26E+13 & 28.0\% & 5.8\% \\
(a) Open & 10.0 & 8.0 & 3.69E+13 & 45.7\% & 5.9\% \\
(a) Open & 10.0 & 10.0 & 4.91E+13 & 60.8\% & 5.3\% \\

\hline
 
(b) Closed & 5.0 & 6.0 & 2.40E+13 & 29.7\% & 27.4\% \\
(b) Closed & 5.0 & 8.0 & 2.83E+13 & 35.1\% & 35.4\% \\
(b) Closed & 5.0 & 10.0 & 2.94E+13 & 36.7\% & 39.1\% \\

\hline

(c) Diamond & 10.0 & 10.0 & 2.50E+13 & 31.0\% & 6.3\% \\

\hline

(d) Ellipse & 10.0 & 10.0 & 2.05E+13 & 25.4\% & 3.9\% \\
\hline
\hline
\end{tabular}
\label{table:1}
\end{table}

\subsection{Particle flow method of determining ABEP intake performance}
A second method to perform the intake performance tests is less conventional than the previous one. It employs the use of quartz crystal micro-balances (QCM) as a gas flow sensor. These are versatile sensors commonly employed in thin film deposition as thickness monitors, although they are now found in many different areas from biosensors to aerospace applications \cite{Qiao2016,Alassi2017,Dirri2019}. QCMs have also been employed as chemical/gas sensors to measure condensation/adsorption of gases with enough sensitivity to detect deposition of atomic layers \cite{Levenson1971,Tsionsky1994,Rahtu2001,Balasingam2021}. 

\begin{figure}[hbt!]
\centering
\includegraphics[width=.7\textwidth]{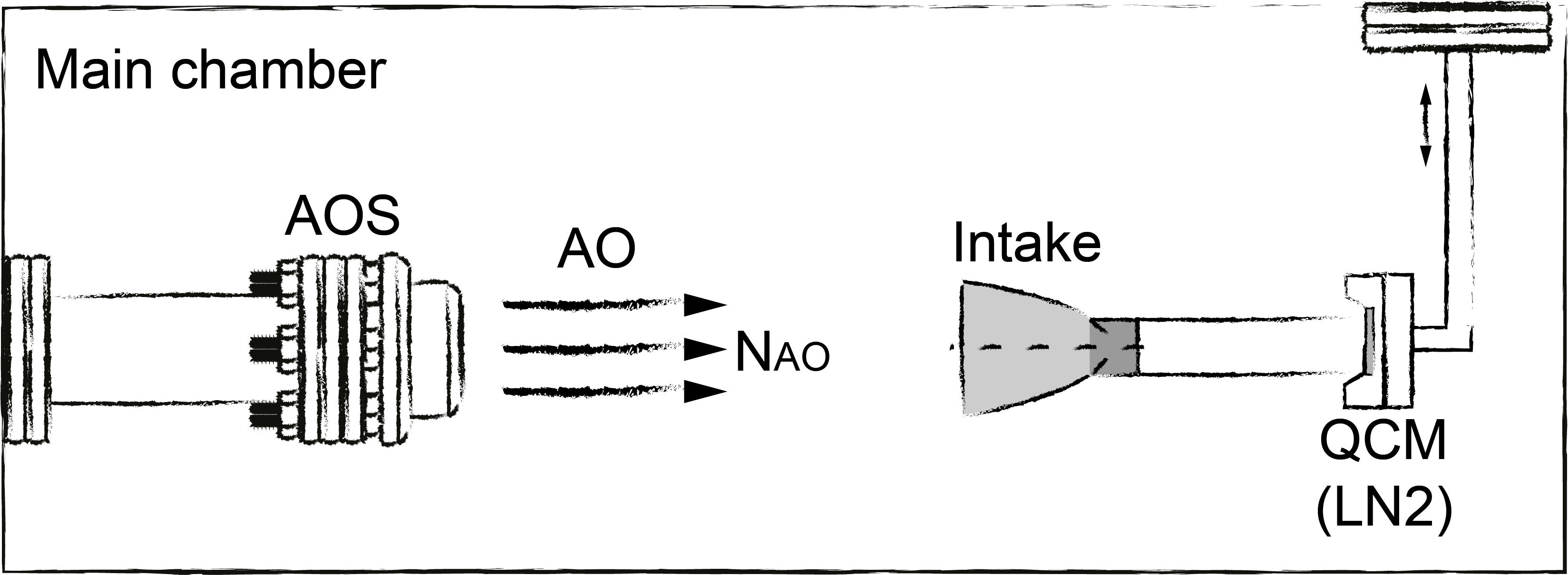}
\caption{Experimental setup for measuring intake performance using a quartz crystal microbalance (QCM) sensor. The crystal is mounted downstream of the intake.} 
\label{fig:QCMSetup}
\end{figure}

Such applications require the surface of the crystal to be coated with a material that erodes under exposure to the atomic oxygen flow at a known rate. Because we are mainly focused on the detection of atomic oxygen, polymer coats such as Kapton are a suitable choice since their erosion rate in an LEO AO environment has been well studied and characterized \cite{Tagawa2003,Banks2006}. The proposed experimental setup for QCM measurements is shown in fig. \ref{fig:QCMSetup}, where the sensor is mounted on a retractable arm that allows for the sensor's position to be adjusted at the outlet of the intake during the tests and maintained at a steady temperature. 

The intake's performance is assessed by the variation of the resonance frequency of the QCM, $f_{0}$. As the oxygen erodes the Kapton coating of the crystal, its mass decreases, and a shift of $f_{0}$ to higher frequencies is observed. This frequency shift can then be related to the incident flux using the empirical erosion rate measurements for the surface material used, and hence the intake's collection efficiency can be determined. Keeping the crystal at a constant temperature during the experiments also contributes to the stability of the measurement and the reduction of systematic errors. An alternative to temperature control is the use of a second crystal as a reference, with the second sensor not being exposed to the back of the intake. Having a second QCM allows for the monitoring of the background and the identification of any changes that could interfere with the experiment. Ideally, the reference crystal would have its temperature separately controlled in an attempt to minimize disturbances in the measurements.

Using outlet flux values determined from the DSMC simulations discussed above, alongside empirical data on erosion yields for a 5.0 eV AO beam, we can estimate the frequency shift expected for different intake characterization experiments. This energy AO beam was considered here since it is within the range of expected ROAR capabilities and is the most well-documented for this process. It is most applicable in the case of a specular surface that results in particles reaching the back of the inlet without significant energy loss through diffuse collisions. By substituting the QCM with the INMS at the outlet of the intake, the particle energy distribution, especially for non-specular surfaces, can be determined empirically and used for erosion rate calibration, compared to the 5.0 eV benchmark derived from the literature. 

Table \ref{table:2} shows the frequency shift per second calculated for intakes of 4 levels of specular\/diffuse reflection, at $0^\circ$ and $10^\circ$ incidence angles. These results assume a $3.0\times10^{-24} \text{cm}^3$ atom$^{-1}$ erosion rate, and a QCM with baseline resonant frequency $f_0$ of 5Mhz coated in Kapton-H \cite{Banks2006}. It also assumes the QCM covers the outlet of the intake, and that particles scattered off the crystal do not contribute to secondary erosion. The frequency shift rate $\dot{f}$ was calculated using:

\begin{align}
    \label{frequencyShiftEq}
    \dot{f} &= -2.26\times10^{-6} (\dot{m} f_0^2) / A
\end{align}

where $\dot{f}$ is the frequency shift rate, $\dot{m}$ is the mass loss rate and $A$ is the surface area of the QCM \cite{Tagawa2003}. Table \ref{table:2} shows that a frequency shift rate of up to $0.247$ Hz s$^{-1}$ is expected in the collection efficiency case of a fully specular intake with the incoming flow at $0^\circ$ incidence, which is comfortably within the detection range of QCM systems for an exposure time on the order of minutes. At the lower end, for a fully diffuse surface with a beam at $10^\circ$ angle of incidence, the frequency shift rate drops by an order of magnitude to $0.015$ Hz s$^{-1}$, which would easily allow for these different conditions to be identified.

\begin{table}[hbt!]
\caption{\label{table:2}Expected frequency shift rate for QCM positioned at intake outlet. }
\centering
\begin{tabular}{c c c}
\hline
\hline  & $0^\circ$ incidence & $10^\circ$ incidence \\\cline{2-2}\cline{3-2}
Surface Type & $\dot{f}$, Hz s$^{-1}$&  $\dot{f}$, Hz s$^{-1}$ \\
\hline
100\% specular & 0.247 & 0.215 \\
80\% specular & 0.140 & 0.085 \\
50\% specular & 0.078 & 0.064 \\
diffuse & 0.033 & 0.015 \\
\hline
\hline
\end{tabular}
\end{table}

\section{Challenges of characterizing ABEP intakes in ROAR}

The characterization of the intake's performance in an AO exposure facility such as ROAR is to date a completely new approach to the development of technologies for VLEO. As the facility was designed to simulate the atmospheric conditions that are equivalent to that of the real environment, in terms of the flux, kinetic energy, and flow regime of AO atoms (free molecular flow), it offers a unique opportunity for testing payloads and materials for space applications at very low Earth orbits. 

Although the benefits of performing tests in ROAR are many, there are a few challenges intrinsic to the measurements that are not exclusive of ROAR, the first being the size of the AO beam. As mentioned, the AO beam is expected to be around $1$ \si{cm} in diameter, which imposes a limit on the dimensions of the testing pieces. Such beam sizes are appropriate for material testing as samples of $1$ \si{cm^{2}} are of adequate sizes for many surface characterization techniques. For intake tests, on the other hand, these dimensions are a significant constraint as it means that only sub-scaled designs can be analyzed.

The restriction in dimensions of the sub-scale designs considerably affects the characterization of their performance and poses some interesting challenges. When considering the pressure measurements, for example, an AO flux of $10^{15}$ \si{atom} \si{cm^{-2} s^{-1}} and collection efficiency of $92\%$, the pressure at the thruster discharge channel would be of the order of $2\cdot10^{-2}$ to $5\cdot10^{-1}$ \si{hPa} for channel diameters of $1$-$5$ \si{mm}. These pressure values for the diameters considered mean a change from free molecular to transitional flow, and such variations in conditions must be observed. To overcome this, the thruster discharge channel can be altered, with the consequent effects of that having to be taken into account from both the theoretical and experimental perspectives. Similar concerns apply to the QCM method as well, and likewise, a geometry optimization of the intake may be required for measurement, but not for the collection of particles as this has been already performed.

Another challenge to be considered is the design nature of most of the currently available AO exposure facilities. Unlike ROAR, the majority of them are dedicated experiments, designed and constructed to efficiently characterize the erosion induced by AO, rather than the aerodynamic performance of a structure. This means they offer limited insight on the gas-surface interaction regime taking place, and how that impacts intake performance. Furthermore, access to these experiments and flexibility regarding the type of measurements that can be performed could at times be considered limited.  

Since ROAR was designed to provide many different access ports to the system, it possesses a level of adaptability that tries to overcome this issue. This feature implies that, firstly, the methodologies considered in this article do not correspond to the only ways of characterizing intakes in a ground-based AO exposure facility; different techniques can be employed depending ultimately on the type of experiments they are aimed for. Secondly, it also means that the experiment itself can be used to perform tests other than the characterization of materials and intakes for VLEO applications. For instance, ROAR can be used also for the development of new sensors and technologies for payloads that are designed and built to operate in conditions where the interaction with atomic oxygen is relevant.

\section{Conclusion}

This article presented a description of the Rarefied Orbital Aerodynamics Research (ROAR) facility currently being commissioned at The University of Manchester, and its utility for oxygen exposure and intake performance analyses. It is designed to characterize the gas-surface interactions between the beam of hyperthermal oxygen atoms and the surface of different materials in conditions that are similar to those found in the real environment, \textit{i.e.}, free molecular flow and AO flux that are representative of very low Earth orbits. The aim of studying these interactions is to identify materials that can reflect oxygen atoms specularly, and by doing so generate less drag in the process. These materials can be applied to the development of new concepts and technologies for VLEO, such as the optimization of intake designs for atmosphere-breathing electric propulsion systems. 

Here are presented and discussed two different methods to characterize the ABEP intake's performance in ROAR. The first approach is to measure the pressure differences between the extremities of the intake. The pumping speed at the end of the intake can be adjusted by a conductance control valve and the auxiliary chamber can be completely isolated from the rest of the experiment by a valve. With this configuration, it is possible to determine the pumping speed, the conductance of the connections between the chambers, and consequently the pressure increase due to the intake's collection. From the pressure measurements and knowledge of the system's dimensions and incoming flow, the number of particles collected can be calculated, which leads consequently to the determination of the collection efficiency.

The second method suggested is the use of a gas sensor to monitor the number of particles going through the intake. Quartz crystal microbalances have been used as such sensors with resolutions that are high enough to detect the erosion of atomic layers. Since the intake will be tested against oxygen, the crystal could be covered in a polymer material like Kapton-H, since its erosion rate under exposure to AO is well characterized. The expected frequency shift rate of such a system would allow for an independent method from pressure measurements to characterize surface interaction regimes or geometric collection efficiencies. 

The challenges and benefits of performing such tests in an AO exposure facility similar to ROAR have also been addressed. The most attractive benefit is the type of experimental data this type of experiment can provide given that it is designed to study the gas-surface interactions instead of erosion mechanisms. Among the limitations is the dimension of the AO beam that constrains the sizes of intakes that can be characterized, access to experiments, and their adaptability to accommodate different types of payloads and sensors.

It is important to emphasize that the methods, as well as ROAR's possible applications in the study and development of new materials and technologies for VLEO, are not limited to the ones considered in this work. The experiment represents a new tool and platform to assist in the efforts of enabling very low Earth orbit technologies for future missions and applications. 

\section*{Funding Sources}
This project has received funding from the European Union’s Horizon 2020 research and innovation program under grant agreement No. $737183$. This reflects only the author’s view and the European Commission is not responsible for any use that may be made of the information it contains. This work is also supported by the UK Space Agency Space Placement in Industry program.

\bibliographystyle{unsrt}  
\bibliography{sample}

\begin{thebibliography}{10}

\bibitem{Visentine1989}
J.~T. Visentine.
\newblock Environmental definition of the earth's neutral atmosphere.
\newblock Technical Report N89-23539, NASA Lyndon B. Johnson Space Center, 1989.

\bibitem{Murad1996}
Edmond Murad.
\newblock Spacecraft interaction with atmospheric species in low earth orbit.
\newblock {\em Journal of Spacecraft and Rockets}, 33(1):131--136, 1996.

\bibitem{Zhang2002}
Jianming Zhang, Donna~J. Garton, and Timothy~K. Minton.
\newblock Reactive and inelastic scattering dynamics of hyperthermal oxygen atoms on a saturated hydrocarbon surface.
\newblock {\em The Journal of Chemical Physics}, 117(13):6239--6251, 2002.

\bibitem{Reddy1995}
M.~Raja Reddy.
\newblock Effect of low earth orbit atomic oxygen on spacecraft materials.
\newblock {\em Journal of Materials Science}, 30(2):281--307, Jan 1995.

\bibitem{Banks2003}
Bruce~A. Banks, Sharon~K.R. Miller, Kim~K. de~Groh, and Rikako Demko.
\newblock Atomic oxygen effects on spacecraft materials.
\newblock Technical Report NASA/TM--2003-212484, National Aeronautics and Space Administration Glenn Research Center Cleveland State University, https://ntrs.nasa.gov/search.jsp?R=20030062195 2017-11-01T10:19:06+00:00Z, June 2003.

\bibitem{Banks2004}
Bruce~A. Banks, Kim~K. de~Groh, and Sharon~K. Miller.
\newblock Low earth orbital atomic oxygen interactions with spacecraft materials.
\newblock Technical Report NASA/TM—2004-213400, Glenn Research Center, Cleveland, Ohio, 2004.

\bibitem{MoeMoeWallace1998}
Kenneth Moe, Mildred~M. Moe, and Steven~D. Wallace.
\newblock Improved satellite drag coefficient calculations from orbital measurements of energy accommodation.
\newblock {\em Journal of Spacecraft and Rockets}, 35(3):266--272, 1998.

\bibitem{Chernik2009}
V.~N. Chernik, L.~S. Novikov, and T.~N. Smirnova.
\newblock Ground‐based atomic oxygen tests of pristine and protected polymeric threads.
\newblock {\em AIP Conference Proceedings}, 1087(1):107--116, 2009.

\bibitem{Roberts2019}
P.~Roberts, N.~Crips, V.~T.~A. Oiko, S.~Edmondson, S.~Haigh, C.~Huyton, S.~Livadiotti, R.~Lyons, K.~Smith, L.~Sinpetru, A.~Straker, and S.~Worral.
\newblock Discoverer – making commercial satellite operations in very low earth orbit a reality.
\newblock International Astronautical Congress (IAC), 70th International Astronautical Congress, October 2019.

\bibitem{TAGAWASEIKYUMAEDAEtAl2006}
Masahito Tagawa, Shinsuke Seikyu, Ken-ichi Maeda, Kumiko Yokota, and Nobuo Ohmae.
\newblock Effect of surface charging on the erosion rate of polyimide under 5 ev atomic oxygen beam exposure.
\newblock In Jacob~I. Kleiman, editor, {\em Protection of Materials and Structures from the Space Environment}, pages 51--59, Dordrecht, 2006. Springer Netherlands.

\bibitem{MurrayPilinskiSmollEtAl2017}
Vanessa~J. Murray, Marcin~D. Pilinski, Eric~J. Smoll, Min Qian, Timothy~K. Minton, Stojan~M. Madzunkov, and Murray~R. Darrach.
\newblock Gas–surface scattering dynamics applied to concentration of gases for mass spectrometry in tenuous atmospheres.
\newblock {\em The Journal of Physical Chemistry C}, 121(14):7903--7922, 2017.

\bibitem{YokotaTagawaOhmae2003}
Kumiko Yokota, Masahito Tagawa, and Nobuo Ohmae.
\newblock Temperature dependence in erosion rates of polyimide under hyperthermal atomic oxygen exposures.
\newblock {\em Journal of Spacecraft and Rockets}, 40(1):143--144, 2003.

\bibitem{BuczalaBrunsvoldMinton2006}
Deanna~M. Buczala, Amy~L. Brunsvold, and Timothy~K. Minton.
\newblock Erosion of kapton h® by hyperthermal atomic oxygen.
\newblock {\em Journal of Spacecraft and Rockets}, 43(2):421--425, 2006.

\bibitem{ReddySrinivasamurthyAgrawal1993}
M.Raja Reddy, N.~Srinivasamurthy, and B.L. Agrawal.
\newblock Atomic oxygen protective coatings for kapton film: a review.
\newblock {\em Surface and Coatings Technology}, 58(1):1 -- 17, 1993.

\bibitem{Packirisamy1995}
S.~Packirisamy, D.~Schwam, and M.~H. Litt.
\newblock Atomic oxygen resistant coatings for low earth orbit space structures.
\newblock {\em Journal of Materials Science}, 30(2):308--320, 1995.

\bibitem{Shpilman2008}
Z.~Shpilman, I.~Gouzman, G.~Lempert, E.~Grossman, and A.~Hoffman.
\newblock rf plasma system as an atomic oxygen exposure facility.
\newblock {\em Review of Scientific Instruments}, 79(2):025106, 2008.

\bibitem{Goodman1967}
F.O. Goodman.
\newblock Three-dimensional hard spheres theory of scattering of gas atoms from a solid surface i. limit of large incident speed.
\newblock {\em Surface Science}, 7(3):391 -- 421, 1967.

\bibitem{SomorjaiBrumbach1973}
G.~A. Somorjai and S.~B. Brumbach.
\newblock The interaction of molecular beams with solid surfaces.
\newblock {\em C R C Critical Reviews in Solid State Sciences}, 4(1-4):429--454, 1973.

\bibitem{Goodman1977}
Frank~O. Goodman.
\newblock Scattering of atoms and molecules by solid surfaces.
\newblock {\em Critical Reviews in Solid State and Materials Sciences}, 7(1):33--80, 1977.

\bibitem{Livadiotti2020}
Sabrina Livadiotti, Nicholas~H. Crisp, Peter~C.E. Roberts, Stephen~D. Worrall, Vitor~T.A. Oiko, Steve Edmondson, Sarah~J. Haigh, Claire Huyton, Katharine~L. Smith, Luciana~A. Sinpetru, Brandon~E.A. Holmes, Jonathan Becedas, Rosa~María Domínguez, Valentín Cañas, Simon Christensen, Anders Mølgaard, Jens Nielsen, Morten Bisgaard, Yung-An Chan, Georg~H. Herdrich, Francesco Romano, Stefanos Fasoulas, Constantin Traub, Daniel Garcia-Almiñana, Silvia Rodriguez-Donaire, Miquel Sureda, Dhiren Kataria, Badia Belkouchi, Alexis Conte, Jose~Santiago Perez, Rachel Villain, and Ron Outlaw.
\newblock A review of gas-surface interaction models for orbital aerodynamics applications.
\newblock {\em Progress in Aerospace Sciences}, 119:100675, 2020.

\bibitem{Oiko2020}
Vitor T.~A. Oiko, Peter C.~E. Roberts, Alejandro Macario-Rojas, Steve Edmondson, Sarah~J. Haigh, Brandon~E.A. Holmes, Sabrina Livadiotti, Nicholas~H. Crisp, Katharine~L. Smith, Luciana~A. Sinpetru, Jonathan Becedas, Rosa~María Domínguez, Valeria Sulliotti-Linner, Simon Christensen, Thomas~Kauffman Jensen, Jens Nielsen, Morten Bisgaard, Yung-An Chan, Georg~H. Herdrich, Francesco Romano, Stefanos Fasoulas, Constantin Traub, Daniel Garcia-Almiñana, Marina Garcia-Berenguer, Silvia Rodriguez-Donaire, Miquel Sureda, Dhiren Kataria, Badia Belkouchi, Alexis Conte, Simon Seminari, and Rachel Villain.
\newblock Ground-based experimental facility for orbital aerodynamics research: design, construction, and characterisation.
\newblock Virtual conference, October 2020. International Astronautical Federation (IAF), Proceedings of the International Astronautical Congress 2020.

\bibitem{Outlaw1987}
R.A. Outlaw, Gar~B. Hoflund, and Gregory~R. Corallo.
\newblock Electron-stimulated desorption of atomic oxygen from polycrystalline ag.
\newblock {\em Applied Surface Science}, 28(3):235--246, 1987.

\bibitem{Outlaw1994}
R.~A. Outlaw and Mark~R. Davidson.
\newblock Small ultrahigh vacuum compatible hyperthermal oxygen atom generator.
\newblock {\em Journal of Vacuum Science \& Technology A: Vacuum, Surfaces, and Films}, 12(3):854--860, 1994.

\bibitem{Outlaw1990}
R.~A. Outlaw.
\newblock O2 and co2 glow‐discharge‐assisted oxygen transport through ag.
\newblock {\em Journal of Applied Physics}, 68(3):1002--1004, 1990.

\bibitem{Wu1993}
D.~Wu, R.~A. Outlaw, and R.~L. Ash.
\newblock Glow‐discharge enhanced permeation of oxygen through silver.
\newblock {\em Journal of Applied Physics}, 74(8):4990--4994, 1993.

\bibitem{Premathilake2019}
Dilshan Premathilake, Ronald~A. Outlaw, Ronald~A. Quinlan, and Charles~E. Byvik.
\newblock Oxygen generation by carbon dioxide glow discharge and separation by permeation through ultrathin silver membranes.
\newblock {\em Earth and Space Science}, 6(4):557--564, 2019.

\bibitem{Attrill2021}
Gemma D.~R. Attrill, Andrew~C. Nicholas, Graham Routledge, Junayd~A. Miah, Dhiren~O. Kataria, Cathryn~N. Mitchell, Robert~J. Watson, James Williams, Alex Agathanggelou, Charles~M. Brown, Scott~A. Budzien, Tobias Carman, Rahil Chaudery, Kenneth~F. Dymond, Ted~T. Finne, Alex Fortnam, Bruce Fritz, Alex Hands, Peter~J. Marquis, Sean Murphy, Talini Pinto-Jayawardena, Duncan Rust, Keith~A. Ryden, Dave Schofield, Andrew~W. Stephan, Kevin Wiggins, and Craig Underwood.
\newblock Coordinated ionospheric reconstruction {CubeSat} experiment ({CIRCE}), in situ and remote ionospheric sensing ({IRIS}) suite.
\newblock {\em Journal of Space Weather and Space Climate}, 11:16, 2021.

\bibitem{Binder2016}
T.~Binder, P.~C. Boldini, F.~Romano, G.~Herdrich, and S.~Fasoulas.
\newblock Transmission probabilities of rarefied flows in the application of atmosphere-breathing electric propulsion.
\newblock {\em AIP Conference Proceedings}, 1786(1):190011, 2016.

\bibitem{Romano2018}
F.~Romano, B.~Massuti-Ballester, T.~Binder, G.~Herdrich, S.~Fasoulas, and T.~Schönherr.
\newblock System analysis and test-bed for an atmosphere-breathing electric propulsion system using an inductive plasma thruster.
\newblock {\em Acta Astronautica}, 147:114--126, 2018.

\bibitem{Romano2020}
F.~Romano, Y.-A. Chan, G.~Herdrich, C.~Traub, S.~Fasoulas, P.C.E. Roberts, K.~Smith, S.~Edmondson, S.~Haigh, N.H. Crisp, V.T.A. Oiko, S.D. Worrall, S.~Livadiotti, C.~Huyton, L.A. Sinpetru, A.~Straker, J.~Becedas, R.M. Domínguez, D.~González, V.~Cañas, V.~Sulliotti-Linner, V.~Hanessian, A.~Mølgaard, J.~Nielsen, M.~Bisgaard, D.~Garcia-Almiñana, S.~Rodriguez-Donaire, M.~Sureda, D.~Kataria, R.~Outlaw, R.~Villain, J.S. Perez, A.~Conte, B.~Belkouchi, A.~Schwalber, and B.~Heißerer.
\newblock Rf helicon-based inductive plasma thruster (ipt) design for an atmosphere-breathing electric propulsion system (abep).
\newblock {\em Acta Astronautica}, 176:476--483, 2020.

\bibitem{Romano2021}
F.~Romano, J.~Espinosa-Orozco, M.~Pfeiffer, G.~Herdrich, N.H. Crisp, P.C.E. Roberts, B.E.A. Holmes, S.~Edmondson, S.~Haigh, S.~Livadiotti, A.~Macario-Rojas, V.T.A. Oiko, L.A. Sinpetru, K.~Smith, J.~Becedas, V.~Sulliotti-Linner, M.~Bisgaard, S.~Christensen, V.~Hanessian, T.~Kauffman Jensen, J.~Nielsen, Y.-A. Chan, S.~Fasoulas, C.~Traub, D.~García-Almiñana, S.~Rodríguez-Donaire, M.~Sureda, D.~Kataria, B.~Belkouchi, A.~Conte, S.~Seminari, and R.~Villain.
\newblock Intake design for an atmosphere-breathing electric propulsion system (abep).
\newblock {\em Acta Astronautica}, 187:225--235, 2021.

\bibitem{Singh2015}
Lake~A. Singh and Mitchell~L.R. Walker.
\newblock A review of research in low earth orbit propellant collection.
\newblock {\em Progress in Aerospace Sciences}, 75:15--25, 2015.

\bibitem{Romano2022}
F~Romano, G~Herdrich, Y.-A. Chan, N~H Crisp, P~C~E Roberts, B~E~A Holmes, S~Edmondson, S~Haigh, A~Macario-Rojas, V~T~A Oiko, L~A Sinpetru, K~Smith, J~Becedas, V~Sulliotti-Linner, M~Bisgaard, S~Christensen, V~Hanessian, T~Kauffman Jensen, J~Nielsen, S~Fasoulas, C~Traub, D~Garc{\'{i}}a-Almi{\~{n}}ana, S~Rodr{\'{i}}guez-Donaire, M~Sureda, D~Kataria, B~Belkouchi, A~Conte, S~Seminari, and R~Villain.
\newblock {Design of an intake and a thruster for an atmosphere-breathing electric propulsion system}.
\newblock {\em CEAS Space Journal}, 14(4):707--715, 2022.

\bibitem{Munz2014}
Claus-Dieter Munz, Monika Auweter-Kurtz, Stefanos Fasoulas, Asim Mirza, Philip Ortwein, Marcel Pfeiffer, and Torsten Stindl.
\newblock Coupled particle-in-cell and direct simulation monte carlo method for simulating reactive plasma flows.
\newblock {\em Comptes Rendus Mécanique}, 342(10):662--670, 2014.
\newblock Theoretical and numerical approaches for Vlasov-maxwell equations.

\bibitem{Fasoulas2019}
S.~Fasoulas, C.-D. Munz, M.~Pfeiffer, J.~Beyer, T.~Binder, S.~Copplestone, A.~Mirza, P.~Nizenkov, P.~Ortwein, and W.~Reschke.
\newblock Combining particle-in-cell and direct simulation monte carlo for the simulation of reactive plasma flows.
\newblock {\em Physics of Fluids}, 31(7):072006, 2019.

\bibitem{Brabants2021}
C~Brabants.
\newblock Development of a methodology for vleo intake performance testing in a low density facility.
\newblock thesis, Faculté des Sciences appliquées, Liège, Belgique, June 2021.

\bibitem{Rapisarda2021}
Claudio Rapisarda.
\newblock {Modelling and simulation of atmosphere-breathing electric propulsion intakes via direct simulation Monte Carlo: A study of the air-breathing ion engine}.
\newblock {\em CEAS Space Journal}, (December), 2021.

\bibitem{Parodi2019}
Pietro Parodi.
\newblock {Analysis and Simulation of an Intake for Air-Breathing Electric Propulsion Systems}.
\newblock (July 2019), 2019.

\bibitem{White2018}
C.~White, M.~K. Borg, T.~J. Scanlon, S.~M. Longshaw, B.~John, D.~R. Emerson, and J.~M. Reese.
\newblock {dsmcFoam+: An OpenFOAM based direct simulation Monte Carlo solver}.
\newblock {\em Computer Physics Communications}, 224:22--43, 2018.

\bibitem{Qiao2016}
Xiaoxi Qiao, Xiangjun Zhang, Yu~Tian, and Yonggang Meng.
\newblock Progresses on the theory and application of quartz crystal microbalance.
\newblock {\em Applied Physics Reviews}, 3(3):031106, 2016.

\bibitem{Alassi2017}
Abdulrahman Alassi, Mohieddine Benammar, and Dan Brett.
\newblock Quartz crystal microbalance electronic interfacing systems: A review.
\newblock {\em Sensors}, 17(12), 2017.

\bibitem{Dirri2019}
Fabrizio Dirri, Ernesto Palomba, Andrea Longobardo, Emiliano Zampetti, Bortolino Saggin, and Diego Scaccabarozzi.
\newblock A review of quartz crystal microbalances for space applications.
\newblock {\em Sensors and Actuators A: Physical}, 287:48--75, 2019.

\bibitem{Levenson1971}
Leonard~L. Levenson.
\newblock Condensation coefficients of argon, krypton, xenon, and carbon dioxide measured with a quartz crystal microbalance.
\newblock {\em Journal of Vacuum Science and Technology}, 8(5):629--635, 1971.

\bibitem{Tsionsky1994}
V.~Tsionsky and E.~Gileadi.
\newblock Use of the quartz crystal microbalance for the study of adsorption from the gas phase.
\newblock {\em Langmuir}, 10(8):2830--2835, 1994.

\bibitem{Rahtu2001}
Antti Rahtu, Teemu Alaranta, and Mikko Ritala.
\newblock In situ quartz crystal microbalance and quadrupole mass spectrometry studies of atomic layer deposition of aluminum oxide from trimethylaluminum and water.
\newblock {\em Langmuir}, 17(21):6506--6509, 2001.

\bibitem{Balasingam2021}
Jenitha~Antony Balasingam, Siddharth Swaminathan, Haleh Nazemi, Calvin Love, Yumna Birjis, and Arezoo Emadi.
\newblock Chemical sensors: Gas sensors, acoustic sensors.
\newblock In {\em Reference Module in Biomedical Sciences}. Elsevier, 2021.

\bibitem{Tagawa2003}
Masahito Tagawa, Kumiko Yokota, Hiroshi Kinoshita, and Nobuo Ohmae.
\newblock {Use of quartz crystal microbalance on the polymer degradation studies regarding atomic oxygen activities in low earth orbit}.
\newblock {\em Proceedings of the 9th International Symposium on Materials in a Space Environmentx}, (540), 2003.

\bibitem{Banks2006}
B.~A. Banks, D.~L. Waters, S.~D. Thorson, K.~K. deGroh, A.~Snyder, and Miller S.
\newblock Comparison of atomic oxygen erosion yields of materials at various energy and impact angles.
\newblock Technical Report NASA/TM--2006-214363, National Aeronautics and Space Administration, https://ntrs.nasa.gov/search.jsp?R=20060047719 2017-11-01T10:32:18+00:00Z, August 2006.

\end{thebibliography}

\end{document}